\documentclass[showpacs,twocolumn,preprintnumbers,amsmath,amssymb,pra]{revtex4}

\usepackage{graphicx}
\usepackage{dcolumn}
\usepackage{bm}

\renewcommand{\vec}[1]{{\mathbf #1}}

\begin{document}

\title{Ballistic quench-induced correlation waves in ultracold gases}

\author{John P. Corson}
\author{John L. Bohn}

\affiliation{JILA, NIST and Department of Physics, University of Colorado, Boulder, Colorado 80309-0440, USA}

\date{\today}

\begin{abstract}
We investigate the wave packet dynamics of a pair of particles that undergoes a rapid change of scattering length. The short-range interactions are modeled in the zero-range limit, where the quench is accomplished by switching the boundary condition of the wave function at vanishing particle separation. This generates a correlation wave that propagates rapidly to nonzero particle separations. We have derived universal, analytic results for this process that lead to a simple phase-space picture of the quench-induced scattering. Intuitively, the strength of the correlation wave relates to the initial contact of the system. We find that, in one spatial dimension, the $k^{-4}$ tail of the momentum distribution contains a ballistic contribution that does not originate from short-range pair correlations, and a similar conclusion can hold in other dimensionalities depending on the quench protocol. We examine the resultant quench-induced transport in an optical lattice in 1D, and a semiclassical treatment is found to give quantitatively accurate estimates for the transport probabilities.

\end{abstract}

\maketitle

\section{Introduction} \label{sec:Introduction}

It is a generic property of wave mechanics that an abrupt change in a system's boundary condition generates waves that propagate outward from the boundary. The tap of a mallet excites phonons in a percussive chime; with a flick of the wrist, a lion tamer snaps his whip; electric pulses in an antenna generate a radio broadcast. A similar phenomenon can occur in ultracold quantum gases, where short-range interactions create an effective boundary condition for the wave function \cite{Bethe1935}. The ``boundary'' occurs at vanishing particle separation, and in the ultracold regime, it is determined by a single parameter called the scattering length, $a$. The scattering length, and hence the boundary condition, can be dynamically tuned by magnetic-field ramps \cite{Claussen2002,Donley2002} or optical switching \cite{Clark2015} in the vicinity of a broad Feshbach resonance \cite{Tiesinga1993,Inouye1998,Courteille1998,Vuletic1999,Chin2010}. In quasi-reduced dimensionalities, where motion is frozen out by tight trapping in one or two dimensions, one can also exploit confinement-induced resonances \cite{Olshanii1998,Haller2010}.

The response of ultracold atoms to a rapid change of scattering length, or ``quench'', is a topic of growing interest in the field of AMO physics. Long-wavelength waves were observed to propagate in the density-density correlation function of a quenched 2D Bose condensate \cite{Hung2013}. In the case of a 3D condensate quenched to resonance, universal dynamics were observed as the system eventually reached an exotic state in which the three-body loss was unexpectedly low \cite{Makotyn2014}. The super-Tonks-Girardeau state \cite{Astrakharchik2005} was created in a 1D Bose gas whose interactions were quenched from strong repulsion to strong attraction \cite{Haller2009}. There has been a thrust of theory work to accompany these exciting experimental advances, both for bosons \cite{Carusotto2010,Gritsev2010,Muth2010,Zill2015,Kormos2014,Natu2013,Sykes2014,Corson2015,Yin2013,Kain2014,Kira2014,Kira2015a,Kira2015b,Yin2016,Piroli2016,Iyer2012,Iyer2013,Rancon2013,Rancon2014} and for fermions \cite{Yuzbashyan2015,Dong2015,Peronaci2015,Yoon2015}. 

Dynamical waves that propagate in the pair correlation function in response to an interaction quench, hereafter referred to as ``quench-induced correlation waves'', have been discussed at length in the context of several many-body models. For the case of 2D and 3D quenched Bose condensates, these correlations have been calculated in the Bogoliubov approximation \cite{Hung2013,Carusotto2010,Natu2013} and with quantum kinetic theory \cite{Kira2015b}. Numerical results were presented for quenched 1D Bose gases in Refs. \cite{Gritsev2010, Muth2010, Zill2015}, with analytical results for the Tonks-Girardeau regime given in Ref. \cite{Kormos2014}. Other studies have calculated the spreading of correlations in quenched single-band Hubbard models using matrix-product-state \cite{Barmettler2012} and variational-Monte-Carlo \cite{Carleo2014} algorithms, accompanying recent experimental progress in that realm \cite{Cheneau2012}. 

Our work takes a different, but complementary, approach to correlation waves. We consider the question: What does an interaction quench (alternatively, a quenched boundary condition) do to the relative wave function for a pair of particles? This question lies at the root of the many-body quench problem, where interactions are pairwise and three-body correlations are often negligible. Two-body models offer the advantage that they can be solved exactly and give direct access to the wave function \cite{Busch1998}. Moreover, they are immediately relevant to few-body systems in optical tweezers \cite{Kaufman2014} and deep optical lattices \cite{Stoferle2006}. In many instances, they have given insight into understanding nonequilibrium many-body phenomena \cite{Boschi2014,Goral2005,Mies2000,Borca2003,Goral2004}. They have also been shown to give a quantitative description of short-time short-range pair correlations in certain nonequilibrium many-body systems \cite{Sykes2014,Corson2015}, a result that is extended in the present study. 

In this paper, we show that two-body models give an intuitive description of the physics behind quench-induced correlation waves. 
Section~\ref{sec: ballistic waves} reviews ballistic expansion from the standpoint of a quench. Our phase-space analysis shows that the correlation waves propagate ballistically, ie. as if they were free particles. We demonstrate that these waves, which are inherently nonlocal, can contribute to the $k^{-4}$ tail of a dynamical momentum distribution. This result is unexpected considering that the ideas surrounding Tan's contact relate the $k^{-4}$ tail exclusively to local correlations \cite{Tan2008a,Tan2008b,Tan2008c}. In Sec.~\ref{sec: arbitrary quenches}, we discuss the leading-order behavior of the momentum distribution for arbitrary interaction quenches. We find that there is generally a competition between short-range and ballistic physics in the large-momentum limit, an effect that is absent in equilibrium scenarios. Additionally, we find that the amplitude of the correlation wave is determined chiefly by the initial and final scattering lengths, and also by the initial amplitude of the wave function at vanishing particle separation. Section \ref{sec: lattice transport} outlines our solution of the two-body quench problem in the presence of an external lattice potential. We show that ballistic correlation waves can propagate even in deep lattices, and we present a simple semiclassical model that yields accurate estimates for the transport that occurs. Section \ref{sec: conclusion} concludes our study.


\section{Ballistic waves} \label{sec: ballistic waves}


It is instructive to begin with the simplest case in which quench-induced correlation waves occur: a measurement of the momentum distribution of a strongly interacting ultracold gas. The general method is to rapidly turn off the external trap and interactions, thereby freezing the momentum distribution of the gas, and then to allow the sample to expand freely before imaging. After expansion, the image represents the column-integrated momentum distribution of the gas. Correlation waves are generated by this simple protocol, as we now demonstrate.



\begin{figure}
\includegraphics[width=0.4\textwidth]{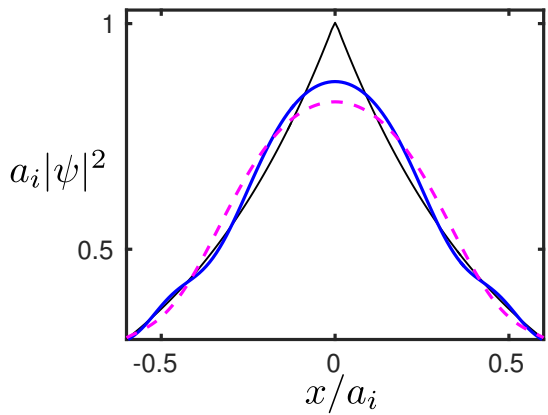}
\caption{(Color online) 
Ballistic expansion of a bound-state wave function. The black (thin) line represents the initial wave function. The blue (thick, solid) line represents the wave function at $\hbar t / 2\mu a_i^2 = 0.008$, and the magenta (dashed) line is the wave function at $\hbar t / 2\mu a_i^2 = 0.015$.
 }
\label{fig: ball wave 1}
\end{figure}

The above-described procedure constitutes an interaction quench in the sense that, trap effects aside, the scattering length is rapidly changed from some initial value ($a_i$) to some final value ($a_f$). The effect on the wave function can be seen in the ballistic expansion of a bound pair of interacting particles in 1D. In terms of the particle separation $x$ and coupling constant $g_{\rm 1D}$, the short-range interaction potential is
\begin{equation} \label{eq: Vint}
V_{\mathrm{int}}(x) = g_{\rm 1D}\delta (x) .
\end{equation}
One can define a 1D scattering length via $a=-\hbar^2/\mu g_{\rm 1D}$, where $\mu$ is the reduced mass for the pair. The interactions are attractive (repulsive) for $a>0$ ($a<0$), and they vanish for $a=\pm \infty$. For an initially bound pair of atoms, the relative wave function is
\begin{equation} \label{eq: bound state}
\psi(x,t=0)=\frac{1}{\sqrt{a_i}}\mathrm{e}^{-|x|/a_i} 
\end{equation}
The time-dependent solution, upon turning off interactions ($a_f\rightarrow \pm\infty$), is most compactly written in momentum space. We define the Fourier transform as
\begin{equation} \label{eq: fourier transform 1d}
\tilde{f}(k)\equiv \int dx \mathrm{e}^{-ikx}f(x),
\end{equation}
and the dynamical wave function is given by
\begin{equation} \label{eq: ballistic expansion}
\tilde{\psi}(k,t)=\frac{2\sqrt{a_i}}{1+k^2 a_i^2}\mathrm{e}^{-iE_kt/\hbar}
\end{equation}
where $E_k = \hbar^2 k^2/2\mu$ is the relative kinetic energy. The short-time dynamics of the position-space wave function is shown in Fig.~\ref{fig: ball wave 1}. At $t=0$, the wave function has a kink at vanishing particle separation. This kink is absent for $t>0$, where we see a correlation wave that propagates to larger particle separations.

\begin{figure}
\includegraphics[width=0.48\textwidth]{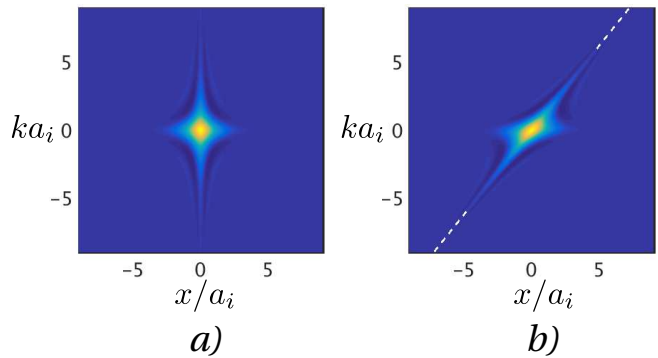}
\caption{(Color online) 
Wigner distribution Eq.~\eqref{eq: wigner dist} for the ballistic expansion of a 1D bound state wave function at (a) $t=0$ and (b) $\hbar t/2\mu a_i^2=0.4 $. The dashed white line in (b) represents the formula $x=2 \hbar k t / m$, which corresponds to the separation of two classical particles that start out on top of each other ($x=0$) and then fly apart with momenta $\pm \hbar k$.
 }
\label{fig: wig1}
\end{figure}

The ballistic expansion dynamics can be easily visualized with a phase-space representation. The Wigner function \cite{Wigner1932}
\begin{equation} \label{eq: wigner dist}
W(x,k,t)=\int dy \mathrm{e}^{iky}\psi^*\left(x+\frac{y}{2},t\right)\psi\left(x-\frac{y}{2},t\right)
\end{equation}
gives an approximate sense of the phase-space distribution of the instantaneous quantum state $\psi(x,t)$. The position and momentum distributions can be found by integrating:
\begin{equation}
\begin{aligned}
& |\psi(x,t)|^2=\int \frac{dk}{2\pi}W(x,k,t) \\&
|\tilde{\psi}(k,t)|^2=\int dx W(x,k,t) 
\end{aligned} .
\end{equation}
Figure \ref{fig: wig1}(a) shows the Wigner function of the bound state at $t=0$. Initially, the $k^{-2}$ tail of the momentum-space wave function is responsible for the kink in the position-space wave function at $x=0$ (cf. Fig.~\ref{fig: ball wave 1}). This is typical for wave functions of 1D systems with short-range interactions. It is generally understood that any state that behaves as $\Psi(x) \approx \Psi(0)(1-|x|/a)$ in the short range should have a contribution
\begin{equation} \label{eq: short range?}
\tilde{\Psi}(k) \sim \frac{2\Psi(0)}{ak^2} +\mathcal{O}(\frac{1}{k^3})
\end{equation}
to the large-momentum limit of the momentum-space wave function \cite{Olshanii2003}. This connection between short-range correlations and large-momentum asymptotics led to the development of universal ``contact'' relations for one-dimensional systems \cite{Olshanii2003,Barth2011}, which were subsequently generalized to three-dimensions \cite{Tan2008a,Tan2008b,Tan2008c,Braaten2008,Braaten2011} and two-dimensions \cite{Tan2005,Combescot2009,Valiente2011,Werner2012a,Werner2012b}.

Figure \ref{fig: wig1}(b) shows that, after the interactions are turned off, the large-momentum components of the wave function propagate outwards to larger particle separations. Although the momentum distribution does not change during the dynamics (cf. Eq.~\eqref{eq: ballistic expansion}), the momentum components eventually separate spatially in a semiclassical sense, with the fastest modes moving the farthest. [In the figure, we see that the spatial wings of the phase-space distribution agree very well with the classical problem in which a pair of particles flies apart with momentum $\pm \hbar k$ (white dashed line), similar to the suggestion of Ref.~\cite{Calabrese2006}.] This mechanism leads to the usual correspondence between the expanded spatial distribution and the initial momentum distribution, as probed by ballistic expansion measurements of interacting systems. We point out that such a mapping would not occur if the interactions were turned off adiabatically or if they were left unchanged; it was necessary to quench the system.

It is interesting that ballistic expansion leads to a momentum distribution whose $k^{-4}$ tail does not correspond to a kink in the short-range wave function. Rather, this tail is responsible for the correlation wave that propagates from the short range to the long range, as evidenced in Fig.~\ref{fig: wig1}. One can alternatively view this correlation wave, and hence the $k^{-4}$ tail in the \emph{dynamical} momentum distribution, to be the result of a rapidly disturbed boundary condition. It can be shown that the interaction potential in Eq.~\eqref{eq: Vint} enforces a log-derivative boundary condition
\begin{equation} \label{eq: log derivative 1d}
\left. \frac{\partial_x \psi}{\psi}\right|_{x\rightarrow 0^+} = -\frac{1}{a} 
\end{equation}
for symmetrized wave functions. The quench from $a_i>0$ to $a_f=\pm \infty$ changes this boundary condition discontinuously, thereby generating a correlation wave. We expect intuitively that such a wave should be generated whenever the quench is diabatic and $a_f\neq a_i$. The strength of the wave should depend on the mismatch between the initial and final boundary conditions. For example, the generated wave should be weak when $a_f \approx a_i$, and it should be strong when $a^{-1}$ changes drastically. We expect also that the large-momentum behavior of the wave function should contain both short-range and ballistic contributions, generalizing Eq.~\eqref{eq: short range?}.

\section{Arbitrary Quenches} \label{sec: arbitrary quenches}

Reference \cite{Corson2015} demonstrated that it is possible to find closed-form solutions to the two-body quench problem in 3D for a broad class of initial wave functions. It was also shown that the short-time, zero-range dynamics depend on only three parameters: the initial scattering length $a_i$, the final scattering length $a_f$, and the initial zero-range behavior of the wave function $r\psi(r,0)\big|_{r\rightarrow 0^+}$. It is natural to suppose that a similar universality should persist in the large-momentum content of the quench-induced ballistic wave, as this wave originates in the short range and is a direct response to the change in boundary condition. We indeed find this to be the case in each dimensionality.

The derivation of the large-momentum limit of the 1D dynamical wave function is given in Appendix \ref{sec: appendix}. In short, one must project the initial wave function onto the complete basis of energy eigenstates satisfying the appropriate log-derivative boundary condition, Eq.~\eqref{eq: log derivative 1d}, and then propagate in time. We find that the large-momentum limit of the wave function is  
\begin{equation} \label{eq: ball wave 1d}
\begin{aligned}
\tilde{\psi}(k,t)=\left(\frac{a_f}{a_i}-1  \right)&\frac{2\psi(0,0)}{(k^2a_f-i|k|)}\mathrm{e}^{-iE_k t/\hbar}\\& +\frac{2 \psi(0,t)}{a_f k^2}
+\mathcal{O}(\frac{1}{k^3})
\end{aligned}\quad \quad (1D)
\end{equation}
for $t>0$. The second term shown here comes from the dynamical kink that appears in the short-range wave function for finite values of $a_f$, as in Eq.~\eqref{eq: short range?} and in accordance with our intuition about the contact \cite{Olshanii2003}. The first term represents the ballistic wave that is generated by the quench, similar to Eq.~\eqref{eq: ballistic expansion}. As a consistency check, it is easy to verify that Eq.~\eqref{eq: ball wave 1d} agrees with Eq.~\eqref{eq: ballistic expansion} in the $a_f\rightarrow \pm\infty$ limit. It is also immediately obvious that the ballistic contribution vanishes in the limit that no quench occurs (ie. $a_f\rightarrow a_i$).

\begin{figure}
\includegraphics[width=0.3\textwidth]{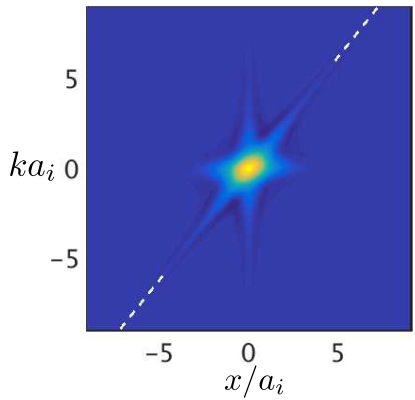}
\caption{(Color online) 
Wigner distribution Eq.~\eqref{eq: wigner dist} for  a 1D bound state wave function quenched to $a_f=2a_i$. The Wigner function is evaluated at the same time as in Fig.~\ref{fig: wig1}(b), and for the same initial scattering length $a_i$. The dashed white line represents the same classical model as shown previously.
 }
\label{fig: wig3}
\end{figure}

It is significant that, after the quench, the large-momentum limit of the wave function has two distinct components that are both $\mathcal{O}(k^{-2})$. This occurs whenever the final scattering length is finite. Figure~\ref{fig: wig3} shows this behavior for the case in which an initial bound state at a scattering length $a_i>0$ is quenched to a final scattering length $a_f=2a_i$. Similar to Fig.~\ref{fig: wig1}(a), we see large-momentum content in the short range that is due to the residual kink in the wave function. Similar to Fig.~\ref{fig: wig1}(b), we see that the quench generates a ballistic correlation wave that rapidly propagates to large particle separations.  This is in strong contrast to equilibrium problems, where only the short-range correlations contribute to the large-momentum asymptotics \cite{Olshanii2003}. For this 1D quench problem, the $k^{-4}$ tail of the one-body momentum distribution ($\sim | \tilde{\psi}(k,t) |^2$) does not correspond perfectly with the zero-range pair probability, indicating that one must exercise care when interpreting the 1D contact in a nonequilibrium context.

\begin{figure}
\includegraphics[width=0.4\textwidth]{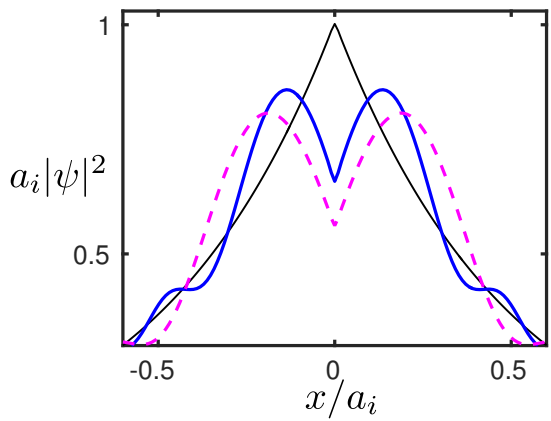}
\caption{(Color online) 
Quenching a bound-state to $a_f=-a_i/2$. The black (thin) line represents the initial wave function. The blue (thick, solid) line represents the wave function at $\hbar t / 2\mu a_i^2 = 0.008$, and the magenta (dashed) line is the wave function at $\hbar t / 2\mu a_i^2 = 0.015$. Compare with Fig.~\ref{fig: ball wave 1}, which depicts a quench to $a_f=\pm\infty$.
 }
\label{fig: ball wave 2}
\end{figure}

The amplitude of the ballistic correlation wave shown in Eq.~\eqref{eq: ball wave 1d} can be adjusted by changing the final scattering length $a_f$. Figure~\ref{fig: ball wave 2} shows the short-time position-space wave function for a bound state that is quenched to $a_f=-a_i/2$, evaluated at the same times as in Fig.~\ref{fig: ball wave 1}. One can see that the quench from attraction ($a_i>0$) to repulsion ($a_f<0$) has increased the amplitude of the correlation wave when compared to the ballistic expansion case ($a_f=\pm\infty$). Equation~\eqref{eq: ball wave 1d} indicates that this enhancement is by approximately a factor of 3.

One may observe from Eq.~\eqref{eq: ball wave 1d} that quenches to the Tonks-Girardeau regime ($a_f\rightarrow 0^-$) generate especially strong ballistic waves. In this limit, the wave function behaves as
\begin{equation} \label{eq: tg limit}
\tilde{\psi}(k,t) = \frac{2\psi(0,0)}{i|k|}\mathrm{e}^{-iE_kt/\hbar}+\mathcal{O}(\frac{1}{k^2})
\end{equation} 
for large $k$ and $t>0$, with the ballistic component dominating the short-range. One can intuit the $k^{-1}$ tail by observing that the final energy of the system is determined by the expectation value of the post-quench Hamiltonian in the initial state; this energy must diverge as $g_f \sim -1/a_f \rightarrow \infty$. For $t>0$, the interaction energy must vanish because $\psi(0,t)=0$. Conservation of energy therefore requires that, after the quench, the kinetic energy diverge:
\begin{equation}
\int \frac{dk}{2\pi}  \left|\tilde{\psi}(k,t)  \right|^2\frac{\hbar^2 k^2}{2\mu} \rightarrow \infty .
\end{equation}
This was first pointed out by the authors of Ref.~\cite{Kormos2014}, who calculated analytically the dynamical density correlations for a many-body system of density $n$ that is quenched from noninteracting to the Tonks-Girardeau regime. Our results connect smoothly with theirs in the short-time ($\hbar n^2 t / m\ll 1$), short-range ($nx\ll 1$) limit. In this limit, their dynamical pair correlations take the form of a relative wave function that behaves exactly as in Eq.~\eqref{eq: tg limit} except that $\psi(0,0)\rightarrow \sqrt{n}$. If we were to simulate the many-body problem with a two-body model, as in Ref.~\cite{Corson2015}, we would use this same prescription. This prescription also leads to quantitative agreement (at short times) with the numerical calculations of $g^{(2)}(0,t)$ in Ref.~\cite{Zill2015}, which considered a broad range of $a_f<0$. This reinforces the idea that properly calibrated few-body models can quantitatively describe short-time short-range correlation phenomena for quenched many-body systems \cite{Corson2015}.


The derivation given in Appendix \ref{sec: appendix} for quenched one-dimensional systems can be straightforwardly generalized to two and three dimensions. In direct analogy with Eq.~\eqref{eq: ball wave 1d}, the results for $t>0$ are
\begin{equation}\label{eq: ball wave 2d}
\begin{aligned}
\tilde{\psi}(\vec k,t)=\ln&\left(\frac{a_i}{a_f}\right)\frac{2\pi\left.\left(\frac{\psi(\rho,0)}{\ln(\rho/b)}  \right)\right|_{\rho\rightarrow 0^+}}{k^2\left(\ln(ka_f)-i\frac{\pi}{2}  \right)}\mathrm{e}^{-iE_k t/\hbar} \\&
-\frac{2\pi\left.\left(\frac{\psi(\rho,t)}{\ln(\rho/b)}\right)\right|_{\rho\rightarrow 0^+}}{k^2} + \mathcal{O}(\frac{1}{k^3})
\end{aligned}\quad (2D)
\end{equation}
where $b>0$ is an arbitrary length scale that makes the argument of the logarithm dimensionless, and
\begin{equation}\label{eq: ball wave 3d}
\begin{aligned}
\tilde{\psi}(\vec k,t)=\bigg(1-&\frac{a_f}{a_i}  \bigg)\frac{4\pi\left.\left(r\psi(r,0)  \right)\right|_{r\rightarrow0^+}}{k^2(1+ika_f)}\mathrm{e}^{-iE_kt/\hbar} \\&
+\frac{4\pi \left.\left(r\psi(r,t)  \right)\right|_{r\rightarrow 0^+}}{k^2}+\mathcal{O}(\frac{1}{k^4})
\end{aligned}\quad (3D) .
\end{equation}
For the two-dimensional case, we define the scattering length with the convention that the bound state has energy $E_B = -\hbar^2/2\mu a^2$ \cite{Mashayekhi2013,endnote1}. Both in 2D and in 3D, we see that the ballistic contribution (first term) vanishes when $a_f=a_i$. It can also be verified that both formulas reduce to the free-particle result when interactions are turned off ($a_f=\infty$ in 2D, and $a_f=0$ in 3D). 

The ballistic contribution in Eqs.~\eqref{eq: ball wave 2d}-\eqref{eq: ball wave 3d} is subleading to the short range in the large-$k$ limit, but it is nevertheless large compared to what one expects in equilibrium. The subleading terms of all equilibrium states are $\mathcal{O}(k^{-4})$ for both dimensionalities. In contrast, we see that the quench induces new subleading structure, which is $\mathcal{O}(k^{-2} \ln^{-1}(k))$ in 2D and $\mathcal{O}(k^{-3})$ in 3D. This subleading behavior in 3D was first pointed out in Ref.~\cite{Sykes2014} for the case of a quench from noninteracting to unitarity, which is a specific instance of Eq.~\eqref{eq: ball wave 3d}. It was also observed in Hartree-Fock-Bogoliubov simulations of quenched Bose-Einstein condensates \cite{Corson2015}, although the nonlocal and ballistic origin of the effect was not obvious in that context.


Despite the subleading nature of the ballistic terms in Eqs.~\eqref{eq: ball wave 2d}-\eqref{eq: ball wave 3d} for finite $a_f$, one can generate leading-order $\mathcal{O}(k^{-2})$ behavior by turning off interactions. This is along the lines of the ballistic-expansion arguments presented in Sec.~\ref{sec: ballistic waves}. If we then turn on interactions before the wave spreads appreciably, the wave function will develop a short-range singularity that will separately contribute a term of $\mathcal{O}(k^{-2})$ to $\tilde{\psi}(k,t)$. As was found for a single quench in 1D [Eq.~\eqref{eq: ball wave 1d}], the short-range and ballistic components can therefore occur at the same order in the large-$k$ limit of the wave function. Again, we conclude that the considerations that relate the momentum tail exclusively to short-range correlations in equilibrium do not always hold outside of equilibrium.


We conclude this section by remarking on the limitations of our ballistic analysis. The zero-range approximation, wherein interactions are represented with boundary conditions at vanishing particle separation, has been enormously successful in describing ultracold quantum gases near broad Feshbach resonances \cite{Chin2010}. This approximation is only valid for momenta satisfying $k r_0 \ll 1$, where $r_0$ is the range of the interaction. In experiments that use tight optical trapping to create quasi-low-dimensional geometries, the oscillator length of the tight trap represents another scale that bounds the ``range'' of the interaction in the reduced dimensionality. The immediate result is that the momentum tails discussed in the context of zero-range models do not extend out indefinitely to large $k$, although the point of breakdown ($k r_0 \sim 1$) might not be easily observable in typical signal to noise by a broad resonance (cf. Ref.~\cite{Stewart2010}). 

Our analysis also invoked the sudden approximation, wherein the scattering length (alternatively, the boundary condition) is assumed to change instantaneously. The consequence is that ballistic modes of arbitrarily large energy are generated by the quench, as shown in Eqs.~\eqref{eq: ball wave 1d}-\eqref{eq: ball wave 3d}. Any experimental realization of the quench protocol will occur over a finite timescale \cite{Claussen2002,Hung2013,Makotyn2014}, and this will lead to an  energy cutoff in the ballistic modes that can be generated. However, optical switching of interactions was demonstrated to be possible on timescales that are short compared to those set by the interaction range \cite{Clark2015}. It follows that experimentally feasible quench times do not introduce any intrinsic constraint on the quench protocol beyond that already introduced by the range of interactions. We note, however, that quenches which are slow compared to the timescale of the interaction range ($\mu r_0^2/\hbar$) should generate correlation waves that are weaker than those produced by diabatic quenches. In the limit that the scattering length is changed adiabatically, no correlation waves are generated.


\section{Lattice Transport} \label{sec: lattice transport}

It makes sense to suppose that the ballistic nature of quench-induced correlation waves should allow for transport over potential-energy barriers. Semiclassically speaking, some part of the $k^{-2}$ ballistic tail in Eq.~\eqref{eq: ball wave 1d} always has enough energy to cross a barrier of finite height. We expect that the amount of transport can be tuned by adjusting the amplitude of the wave and, therefore, the strength of the quench. In this section, we investigate the quench-induced dynamics that occurs for a pair of particles in a single site of a 1D optical lattice. We find that a semiclassical adaptation of our quantum description gives a good measure of the quench-induced transport. 

Interaction-quench effects in an optical lattice were recently discussed in the numerical results of Refs.~\cite{Mistakidis2014,Mistakidis2015}. There, the authors used the multi-layer multi-configuration time-dependent Hartree method for bosons (ML-MCTDHB) to investigate the dynamics of several interacting bosons in a few lattice sites. They found that a quench can trigger rapid transport between wells, as well as breathing and cradle modes within a given well. Such higher-band effects are ignored in typical Hubbard models that only include the lowest Wannier state in the formalism. Bound states, strong interactions, and/or strong quenches may distort the wave function considerably from the Wannier description, thereby necessitating models that encompass higher-bands. The inclusion of higher bands, either in the ML-MCTDHB sense or in the spirit of a multi-band Hubbard model \cite{Dutta2015}, makes it difficult to obtain numerically converged results for many-body systems on a lattice with strong interactions and strong quenches. Our two-body calculation should provide a useful benchmark in quantitatively understanding the rapid transport that takes place after an interaction quench.

The relative and center-of-mass coordinates do not separate for the case of an interacting pair of atoms in an optical lattice, and we therefore resort to numerics to investigate the exact quantum dynamics. Without loss of generality, we consider identical bosons of mass $m$. We simplify expressions by scaling lengths by the lattice spacing $\ell$; we similarly scale energies by $\hbar^2/m\ell^2=2E_R/\pi^2$, where $E_R$ is the recoil energy of the lattice. The time-dependent Schr\"{o}dinger equation for this system can then be written as
\begin{equation}\label{eq: schrod eq 1}
\begin{aligned}
i\frac{\partial \Psi}{\partial t}= -\frac{1}{2}&\frac{\partial^2\Psi}{\partial x_1^2}-\frac{1}{2}\frac{\partial^2 \Psi}{\partial x_2^2}\\&
 + V_{\rm lat}(x_1)\Psi+V_{\rm lat}(x_2)\Psi+V_{\rm int}(x_1-x_2)\Psi
\end{aligned}
\end{equation}
where the interaction potential is given by Eq.~\eqref{eq: Vint}, and the optical lattice potential is given by
\begin{equation}
V_{\rm lat}(x_j)=V_0\sin^2(\pi x_j),
\end{equation}
 and $V_0$ is the depth of the lattice. Inasmuch as the interaction quench directly excites relative momenta, it is convenient to work with the relative coordinate $x=x_1-x_2$ and the center-of-mass coordinate $X=(x_1+x_2)/2$. After  changing variables, one finds that
\begin{equation}\label{eq: schrod eq 2}
\begin{aligned}
i\frac{\partial \Psi}{\partial t} = -&\frac{\partial^2 \Psi}{\partial x^2}-\frac{1}{4}\frac{\partial \Psi}{\partial X^2}\\&
+V_0\left(1-\cos\left(2\pi X\right)\cos\left(\pi x  \right)\right)\Psi+V_{\rm int}(x)\Psi
\end{aligned}.
\end{equation}
The energy eigenstates corresponding to Eqs.~\eqref{eq: schrod eq 1}-\eqref{eq: schrod eq 2} were found numerically in Ref.~\cite{vonStecher2011} (see also Ref.~\cite{Wall2012} for the 3D analogue). Here, we instead solve for the dynamical wave function by time-evolving an initial condition using the split-operator method \cite{Leforestier1991}. We exploit bosonic symmetry [$\Psi(x,X)=\Psi(-x,X)$] by discretizing the wave function only for $x\geq 0$; spectral transforms are taken along $x$ using the discrete cosine transform, and they are taken along $X$ using the fast fourier transform. We model short-range interactions on the spatial grid by employing a potential that has support only at the $x=0$ grid point. We have found that representing $-\frac{2}{a}\delta(x)\rightarrow -\frac{2}{a}\delta_{x,0}/\Delta x$, where $\Delta x$ is the grid spacing along the $x$ direction, leads to the correct log-derivative boundary condition Eq.~\eqref{eq: log derivative 1d} in the limit that $\Delta x \ll |a|$ \cite{endnote3}. 

\begin{figure}
\includegraphics[width=0.45\textwidth]{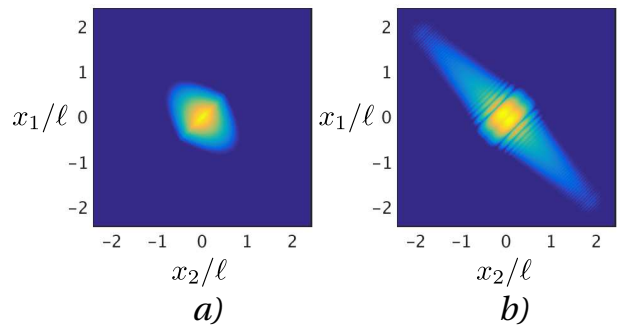}
\caption{(Color online) 
Quenching a bound state from $a_i=0.2\ell$ to $a_f=-a_i$ on a single lattice site. The lattice is assumed to have depth of $10 E_R$. a) The initial two-body probability density $|\Psi(x_1,x_2)|^2$ for a bound state on a lattice site $(t=0)$. b) The same quantity calculated at $\hbar t/m\ell^2=0.01$. Both plots use a logarithmic color scale with the color ranging from a cutoff $10^{-5}$ (purple) to the maximum (yellow). 
 }
\label{fig: lat wave 2}
\end{figure}

Our numerical study focuses on quenched systems for which the induced transport is expected to be the most significant. As indicated in Eq.~\eqref{eq: ball wave 1d}, the amplitude of the ballistic wave is proportional to the initial probability amplitude that the atoms are in the same position, $\psi(0,0)$. This quantity is largest, in equilibrium, when the system is in a bound state. We therefore choose the initial condition for the transport problem Eq.~\eqref{eq: schrod eq 2} to be a bound state in a single lattice site. This configuration represents a subsystem of the state described by Ref.~\cite{Volz2006}, which reported observing a single molecule per lattice site. For a deep lattice, the bottom of the well can be approximated as a harmonic oscillator potential of frequency $\omega=\pi\sqrt{2V_0}$. One can write the approximate initial condition as
\begin{equation} \label{eq: initial condition}
\Psi(x,X)=\psi_0(x)\phi_0(X),
\end{equation}
where $\phi_0(X)=(2/\pi)^{1/4}\mathrm{e}^{-X^2}$ describes the center-of-mass degree of freedom, and $\psi_0(x)$ is the molecular state dressed by the oscillator \cite{Busch1998}. For $a\ll 1/\sqrt{\omega}$, one can show that $\psi_0$ approaches the ordinary bound state given by Eq.~\eqref{eq: bound state}. For our simulations, we will consider an initial bound state of scattering length $a_i=0.2$ in a lattice of depth $V_0=10 E_R$. The two-body probability density associated with this initial condition is shown in separate-particle coordinates in Fig.~\ref{fig: lat wave 2}(a).

As discussed previously, we expect that an interaction quench will generate an energetic correlation wave that propagates over the potential barriers that separate individual lattice sites. This transport is shown in Fig.~\ref{fig: lat wave 2}(b) a short time after quenching to $a_f=-a_i$. The wave has the same general structure as in Fig. \ref{fig: ball wave 2}, with spatially decaying oscillations and a cusp of reduced probability when both particles come together. Even after such a short time, we see that the wave already extends a couple of lattice sites in each direction. 

\begin{figure}
\includegraphics[width=0.45\textwidth]{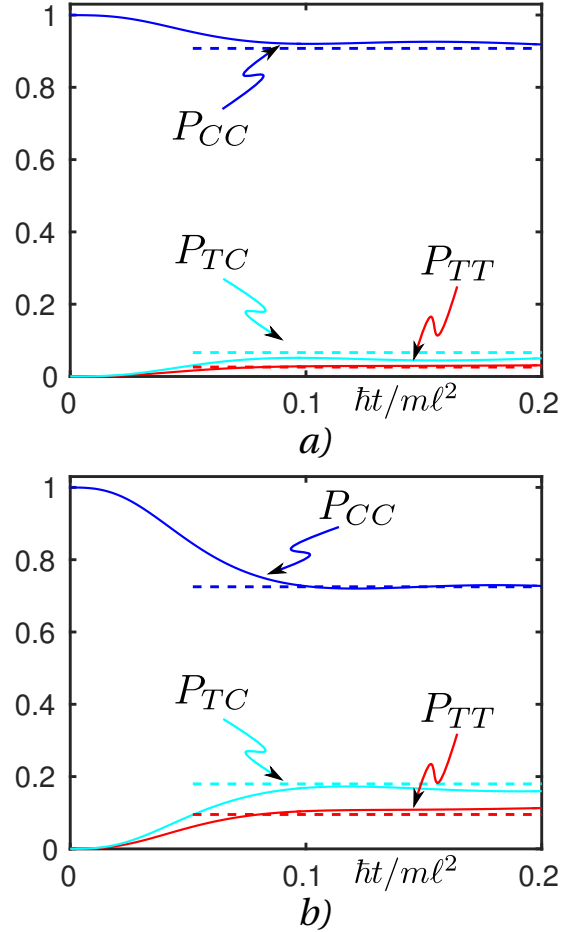}
\caption{(Color online) 
Lattice transport for a bound state that is quenched from $a_i=0.2\ell$ to (a) $a_f=\pm\infty$ and (b) $a_f=-a_i$. The solid blue line denotes the dynamical probability of both atoms occupying the central well of the lattice, $P_{CC}(t)$; the solid cyan line is the probability of a single atom occupying the central well $P_{TC}(t)$; the solid red line is the probability that no atoms occupy the central lattice site, $P_{TT}(t)$. The horizontal dashed lines correspond to the semiclassical estimates for these probabilities.
 }
\label{fig: lat tunneling}
\end{figure}

It is instructive to quantify the amount of quench-induced transport that takes place. We can define a dynamical probability for the liklihood that both atoms remain in the central well:
\begin{equation}
P_{CC}(t)=\int\limits_{|x_1|<\frac{1}{2}}dx_1 \int\limits_{|x_2|<\frac{1}{2}} dx_2
|\Psi(x_1,x_2,t)|^2 .
\end{equation}
In like manner, we also define the probability that both atoms have tunneled,
\begin{equation}
P_{TT}(t) = \int\limits_{|x_1|>\frac{1}{2}}dx_1 \int\limits_{|x_2|>\frac{1}{2}} dx_2
|\Psi(x_1,x_2,t)|^2 ,
\end{equation}
and the probability that a single atom has tunneled,
\begin{equation}
P_{TC}(t)=2\int\limits_{|x_1|>\frac{1}{2}}dx_1 \int\limits_{|x_2|<\frac{1}{2}} dx_2
|\Psi(x_1,x_2,t)|^2 .
\end{equation}
Here, we have exploited the symmetry of the bosonic wave function. The complementarity of the integration regions results in the identity $P_{CC}+P_{TT}+P_{TC}=1$ at all times. These probabilities are plotted for $a_f=\pm\infty$ and $a_f=-a_i$ in Figs.~\ref{fig: lat tunneling}(a) and \ref{fig: lat tunneling}(b), respectively. In both cases, the atoms begin in the central well ($P_{CC}(0)\approx 1$). After the quench, the transport probabilities smoothly saturate to values that depend on $a_f$. We note that the transport is substantial even though the lattice depth is of the order required for a typical Mott-insulating state in 1D \cite{Jaksch1998, Zwerger2003}.

The ballistic description of the previous section leads to an intuitive, semiclassical model of transport. We can estimate the saturated values of $P_{CC}$, $P_{TT}$, and $P_{TC}$ by considering the following question: What fraction of the momentum distribution describes ballistic atoms that are energetic enough to make it over the barrier? 

The simplest analysis can be made for the case in which the interactions are turned off ($a_f=\pm\infty$). Short-range physics then does not contribute to the momentum distribution, and we can consider ballistic effects to stem entirely from the momentum-space version of the initial condition Eq.~\eqref{eq: initial condition}, similar to our analysis in Sec.~\ref{sec: ballistic waves}. One can find the initial two-body wave function $\tilde{\Psi}(k_1,k_2,0)$ from Eq.~\eqref{eq: initial condition} via
\begin{equation} \label{eq: two body momentum dist}
\begin{aligned}
\tilde{\Psi}(k_1,k_2,0)& =\tilde{\psi}_0(k)\tilde{\phi}_0(K) \\&
=\tilde{\psi}_0(k_1-k_2)\tilde{\phi}_0\left(\frac{k_1+k_2}{2}\right)
\end{aligned}
\end{equation}
where we have changed to separate-particle momentum coordinates $k_1$ and $k_2$ from the relative and center-of-mass coordinates $k$ and $K$. In a semiclassical sense, we expect that atoms with kinetic energy $k_i^2/2<V_0$ don't make it over the barrier. Hence, we estimate that the probability for both atoms to stay in the central lattice site is given by
\begin{equation} \label{eq: cc estimate}
P_{CC} \rightarrow \int\limits_{k_1^2/2<V_0} \frac{dk_1}{2\pi} \int\limits_{k_2^2/2<V_0} \frac{dk_2}{2\pi} |\tilde{\Psi}(k_1,k_2,0)|^2.
\end{equation}
Similarly, we estimate the other transport probabilities to be
\begin{equation}\label{eq: transport estimates}
\begin{aligned}
&P_{TT} \rightarrow \int\limits_{k_1^2/2>V_0} \frac{dk_1}{2\pi} \int\limits_{k_2^2/2>V_0} \frac{dk_2}{2\pi} |\tilde{\Psi}(k_1,k_2,0)|^2 \\&
P_{TC} \rightarrow 2 \int\limits_{k_1^2/2>V_0} \frac{dk_1}{2\pi} \int\limits_{k_2^2/2<V_0} \frac{dk_2}{2\pi} |\tilde{\Psi}(k_1,k_2,0)|^2
\end{aligned}.
\end{equation}
These probabilities are plotted as the horizontal dashed lines in Fig.~\ref{fig: lat tunneling}(a), and they agree reasonably well with the saturation observed in the dynamics.

When $a_f$ is finite, the transport estimates should include only the ballistic contribution to the momentum distribution. This much is clear from the fact that, in the absence of a quench, the $k^{-4}$ tail of the momentum distribution contributes to the short range instead of to transport. More generally, the momentum distribution has a mixture of short-range and ballistic effects, as shown in Eq.~\eqref{eq: ball wave 1d} to leading order. We have found that, in most cases, an accurate estimate of transport probabilities requires going beyond leading order so as to suitably include momenta $k \sim \sqrt{2V_0}$. In using Eqs.~\eqref{eq: two body momentum dist} and \eqref{eq: transport estimates}, we replace $\tilde{\psi}_0(k)$ with the full ballistic wave function $\tilde{\psi}_{\rm bal}^{(S)}(k,t)$ derived in Appendix~\ref{sec: appendix} and given by Eq.~\eqref{eq: ball formula 1} \cite{endnote4}. The resulting estimates for the case of $a_f=-a_i$ are plotted as dashed lines in Fig.~\ref{fig: lat tunneling}(b). The increased transport that occurs for this quench is well described by the semiclassical estimate. This agreement owes itself to the fact that a wave of energy $E_k$ incident on a potential barrier of height $V_0$ has near unity transmission for $E_k\gg V_0$. These waves dominate the integrals in Eq.~\eqref{eq: transport estimates} when the quench is strong.

It is interesting that the saturation timescale in Figs.~\ref{fig: lat tunneling}(a)-(b) does not appear to depend on the final scattering length of the quench. We have found that the saturation time is well approximated by the time it takes an atom of momentum $k=\sqrt{2V_0}$ to travel one lattice spacing. This supports our semiclassical description of quench-induced transport. For the lattice depth used in our simulations, the saturation timescale is smaller than the lowest-band tunneling time by more than two orders of magnitude. The higher-band physics at play in this transport process comes from our use of strong interactions \cite{Kohl2005, Diener2006} as well as from the quench itself \cite{Mistakidis2014,Mistakidis2015,Sakmann2011}.

\section{Conclusion} \label{sec: conclusion}

To summarize, we have taken a wave-function-based approach to describe the correlation waves induced by an interaction quench. Our calculations made use of the zero-range approximation for particle-particle interactions, represented here with a scattering-length-dependent boundary condition at vanishing particle separation. Within this approximation, the interaction quench disturbs the boundary and generates a wave that propagates ballistically to nonzero particle separations. We have derived the leading-order behavior of this wave in momentum space for one, two, and three spatial dimensions. These results are intuitive in that the amplitude of the correlation wave depends only on the initial amplitude at the boundary and the scattering length before and after the quench. In each dimensionality, the ballistic contribution to the wave function dominates the next-to-leading-order terms that occur in equilibrium systems. Particularly interesting is the fact that, in one dimension, the $k^{-2}$ tail of the momentum-space wave function is generally determined by both short-range and ballistic effects. Similar results can occur in two and three dimensions, depending on the quench sequence. It is significant that a protocol as simple as a quench can surprise the intuition that usually associates large-momentum behavior exclusively with short-range physics. On this account, our two-body calculations indicate that one must exercise care when interpreting the contact out of equilibrium.



Our simulations reveal that quench-induced correlation waves can cause considerable transport in a 1D optical lattice. The amount of transport that takes place is readily tunable by altering the initial short-range pair probability of the state, as well as the strength of the quench. Our analytic two-body calculation makes possible a semiclassical framework within which both the transport and the saturation time can be estimated with surprising accuracy. We expect that similar results hold for optical-lattice systems in higher dimensionalities whose numerical calculations are more challenging. It would be interesting to see what role these ballistic dynamics might play in a quenched many-body system. For example, the system described in Ref.~\cite{Volz2006}, which was essentially a Mott insulator of molecules in a lattice, might have phase coherence partially restored by the colliding ballistic waves that a quench might generate. One can expect generally that ballistic waves should be damped by collisions in a many-body system. This damping is difficult to model quantitatively without introducing a certain amount of arbitrariness to the theory \cite{Rancon2013,Rancon2014,Kira2014,Kira2015a,Kira2015b}. At the same time, it is the crux of the question of how isolated quantum many-body systems equilibrate despite the high level of excitation provided by a quench. It may be possible to shed light on the matter by investigating how ballistic waves collide even at the few-body level. This remains for future work.

\section*{Acknowledgement}
J.P.C and J.L.B recognize support from the NDSEG fellowship program and the NSF, respectively. We acknowledge helpful conversations with K. R. A. Hazzard.

\appendix

\section{Two-Body Solution in 1D} \label{sec: appendix}

In this appendix, we focus on the two-body solution for interacting particles in one dimension. The solutions for other dimensionalities can be found in essentially the same manner. We simplify expressions by scaling distances by an arbitrary length scale $\xi$ and energies by $\hbar^2/2\mu\xi^2$. In free space, the time-dependent Schr\"{o}dinger equation for the relative wave function $\psi(x,t)$ is
\begin{equation} \label{eq: schrod eq}
i \frac{\partial \psi(x,t)}{\partial t} = -\frac{\partial^2 \psi(x,t)}{\partial x^2}-\frac{2}{a_f}\delta(x)\psi(x,t)
\end{equation}
where $a_f$ is the 1D scattering length after the quench. Without loss of generality, we will consider the particles to be identical bosons. Symmetrization then requires that the relative wave function satisfy $\psi(x,t) = \psi(-x,t)$. The overall effect of the interaction is to enforce the log-derivative boundary condition shown in Eq.~\eqref{eq: log derivative 1d}.

We can propagate a given initial condition $\psi(x,0)$ in time by expanding in the energy eigenstates that satisfy the post-quench log-derivative boundary condition. The scattering states are
\begin{equation} \label{eq: scat eigenstates}
 \psi_{k'}^{(S)}(x)=A_{k'}\left[\sin (k'|x|)-k'a_f \cos(k'x)\right] ,\quad E_{k'}=k'^2 ,
\end{equation}
where 
\begin{equation}
A_{k'} = \frac{1}{\sqrt{2\pi k'(1+k'^2a_f^2)}}
\end{equation}
is a constant that enforces energy normalization. These states are uniquely defined for $k'>0$. For $a_f>0$, the bound state solution is given by Eq.~\eqref{eq: bound state}, with $a_i\rightarrow a_f$, and its energy is $E_B = -1/a_f^2$. As we are focusing on quench-induced scattering, it will be helpful to decompose the time-dependent wave function onto its scattering and bound contributions \cite{Merzbacher}:
\begin{equation}
\psi(x,t)=\psi^{(S)}(x,t)+\psi^{(B)}(x,t) ,
\end{equation}
where
\begin{equation}\label{eq: scat cont}
\psi^{(S)}(x,t)\equiv \int\limits_0^\infty dE_{k'} \mathrm{e}^{-iE_{k'}t}\psi_{k'}^{(S)}(x)\left\langle \left. \psi_{k'}^{(S)}(x')\right| \psi(x',0) \right\rangle
\end{equation}
and
\begin{equation}
\psi^{(B)}(x,t)\equiv \Theta(a_f)\mathrm{e}^{-iE_B t}\psi_B(x)\left\langle \left.\psi_B(x')\right| \psi(x',0)\right\rangle,
\end{equation}
and where $\langle \cdot | \cdot \rangle$ denotes a projection integral. The heaviside function $\Theta(a_f)$ determines whether or not the bound state should be included in the dynamics. We will assume that $\psi(x,0)$ is normalizable and smooth everywhere except possibly for a nontrivial log-derivative at $x=0$.

It is most convenient to solve for the momentum-space wave function, as in Eq.~\eqref{eq: ballistic expansion}. This requires taking the Fourier transform of the energy eigenstates. For $a_f>0$, the Fourier transform of the bound state can be inferred from Eq.~\eqref{eq: ballistic expansion}
\begin{equation}
\tilde{\psi}_B(k)=\frac{2\sqrt{a_f}}{1+k^2a_f^2} .
\end{equation}
The Fourier transform of the scattering states takes a more complicated form, but it can be written as
\begin{equation} \label{eq: scat tran}
\begin{aligned}
\tilde{\psi}_{k'}^{(S)}(k)=A_{k'}&\bigg[-2k'\pi\left( i+k'a_f\right)\delta(k'^2-k^2)  \\& \quad\quad\quad\quad +\frac{2 k'}{k'^2-k^2-i\epsilon}  \bigg] 
\end{aligned},
\end{equation}
where we use the convention $\epsilon\rightarrow 0^+$. Written in this way, there are two parts that compose the scattering contribution $\psi^{(S)}(x,t)$ in Eq.~\eqref{eq: scat cont}. The first part can be evaluated trivially in momentum space by exploiting the delta function in Eq.~\eqref{eq: scat tran}.  We write it as follows:
\begin{equation} \label{eq: ball formula 1}
\begin{aligned}
\tilde{\psi}^{(S)}_{\rm bal}(k,t)\equiv -2\pi|k|&A_{|k|}\left(i+|k|a_f\right)\\&
\times \left\langle \left. \psi_{|k|}^{(S)}(x')\right| \psi(x',0) \right\rangle \mathrm{e}^{-iE_kt}
\end{aligned}
\end{equation}
We call this the ``ballistic'' contribution to the wave function due to its free-particle-like $f(k) \mathrm{e}^{-iE_kt}$ behavior, similar to Eq.~\eqref{eq: ballistic expansion}. The second contribution can be written as
\begin{equation}\label{eq: sr wavefunction}
\begin{aligned}
\tilde{\psi}^{(S)}_{\rm sr}(k,t)=\int\limits_0^\infty 2k'dk' & \mathrm{e}^{-ik'^2 t}\left[\frac{2k'A_{k'}}{k'^2-k^2-i\epsilon}\right]\\&
\times \left\langle \left. \psi_{k'}^{(S)}(x')\right| \psi(x',0) \right\rangle
\end{aligned},
\end{equation} 
and we will see that it generally contributes to the short-range part of the wave function. In sum, we can write
\begin{equation}
\tilde{\psi}(k,t)=\tilde{\psi}^{(S)}_{\rm bal}(k,t)+\tilde{\psi}^{(S)}_{\rm sr}(k,t)+\tilde{\psi}^{(B)}(k,t).
\end{equation}
for the full momentum-space wave function.

\begin{figure}
\includegraphics[width=0.35\textwidth]{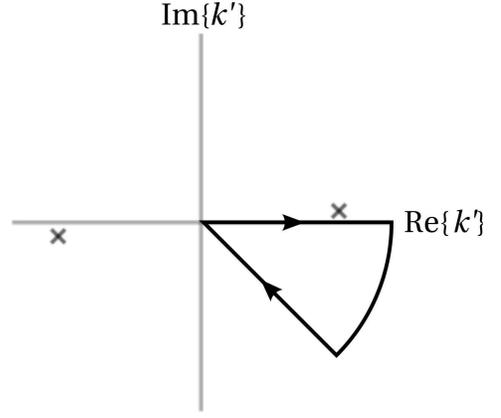}
\caption{
Integration contour in the complex $k'$ plane. The crosses denote the poles of the second term in Eq.~\eqref{eq: scat tran}. The final leg of the integral is along $\mathrm{arg}[k']=-\pi/4$.
 }
\label{fig: contour}
\end{figure}

One can make progress with Eq.~\eqref{eq: sr wavefunction} by exploiting the residue theorem. The integration, as written, is along the positive real $k'$ axis. If we close the contour as shown in Fig.~\ref{fig: contour}, the contribution from the $|k'|\rightarrow\infty$ arc vanishes. The only poles that can contribute residues must come from the analytic continuation of the scattering projection inside the integration loop. We have found empirically that, if $\psi(x,0)$ decays smoothly and without oscillation (such as for a bound state or a ground-state Busch wave function \cite{Busch1998}), the integrand is analytic inside the closed contour and the integral vanishes \cite{endnote2}. The two straight legs of the integral then cancel, and we can rewrite Eq.~\eqref{eq: sr wavefunction} as an integral along $k'=z\mathrm{e}^{-i\frac{\pi}{4}}$ for real, nonnegative $z$:
\begin{equation}\label{eq: sr wave cont}
\begin{aligned}
\tilde{\psi}_{\rm sr}^{(S)}(k,t) & =
-4i\mathrm{e}^{-i\frac{\pi}{4}}\int\limits_0^\infty dz \mathrm{e}^{-z^2 t}\frac{z^2}{-iz^2-k^2} \\&
\quad\quad \times \left[A_{k'}\left\langle \left. \psi_{k'}^{(S)}(x')\right| \psi(x',0) \right\rangle  \right]_{k'\mapsto z\mathrm{e}^{-i\frac{\pi}{4}}}
\\&
=\frac{4i\mathrm{e}^{-i\frac{\pi}{4}}}{k^2}\int\limits_0^\infty dz \frac{\mathrm{e}^{-z^2t}}{1+iz^2/k^2}z^2
\\&
\quad\quad
\times \left[A_{k'}\left\langle \left. \psi_{k'}^{(S)}(x')\right| \psi(x',0) \right\rangle  \right]_{k'\mapsto z\mathrm{e}^{-i\frac{\pi}{4}}}
\end{aligned}
\end{equation}
where the factor in brackets has been analytically continued. This quantity can be evaluated in closed form for several interesting cases, including where $\psi(x,0)$ is an arbitrary bound-state wave function, although the expressions are generally too lengthy to usefully write down. 

The physical significance of $\tilde{\psi}_{\rm sr}^{(S)}(k,t)$ can be seen if one compares it with the zero-range contribution to the scattered wave function, $\psi^{(S)}(0,t)$. Using Eq.~\eqref{eq: scat eigenstates}, we can write the result as
\begin{equation}\label{eq: zr wave}
\begin{aligned}
\psi^{(S)}(0,t)&=\int\limits_0^\infty dE_{k'}\mathrm{e}^{-iE_{k'}t}\psi_{k'}^{(S)}(0) \left\langle \left. \psi_{k'}^{(S)}(x')\right| \psi(x',0) \right\rangle\\&
=-a_f\int\limits_0^\infty dE_{k'}\mathrm{e}^{-iE_{k'}t}k'A_{k'} \left\langle \left. \psi_{k'}^{(S)}(x')\right| \psi(x',0) \right\rangle \\&
=2ia_f\mathrm{e}^{-i\frac{\pi}{4}}\int\limits_0^\infty dz \mathrm{e}^{-z^2t}z^2\\&
\quad\quad
\times\left[A_{k'}\left\langle \left. \psi_{k'}^{(S)}(x')\right| \psi(x',0) \right\rangle  \right]_{k'\mapsto z\mathrm{e}^{-i\frac{\pi}{4}}}
\end{aligned}
\end{equation}
where we have again exploited the integration contour in Fig.~\ref{fig: contour}. A direct comparison of Eq.~\eqref{eq: sr wave cont} and Eq.~\eqref{eq: zr wave} indicates that
\begin{equation} \label{eq: sr wave s}
\tilde{\psi}_{\rm sr}^{(S)}(k,t)=\frac{2\psi^{(S)}(0,t)}{a_fk^2}+\mathcal{O}\left(\frac{1}{k^4}\right)
\end{equation}
for large $k$ satisfying $k^2t\gg 1$. This verifies our claim that $\tilde{\psi}_{\rm sr}^{(S)}$ generally encodes the short-range behavior of the scattered wave. The Gaussian suppression in Eqs.~\eqref{eq: sr wave cont}-\eqref{eq: zr wave} indicates that this contribution to the wave function vanishes in the $t\rightarrow\infty$ limit. This is as expected for an unconfined wave packet composed entirely of scattering states, which must spread out in space as time passes. With significantly less work, one can also show that
\begin{equation} \label{eq: sr wave b}
\tilde{\psi}^{(B)}(k,t)=\frac{2\psi^{(B)}(0,t)}{a_f k^2}+\mathcal{O}\left(\frac{1}{k^4}\right)
\end{equation}
for the bound-state contribution to the dynamical wave function.

We now examine the large-momentum behavior of the ballistic contribution to the wave function, given by Eq.~\eqref{eq: ball formula 1}. It can be shown that
\begin{equation}
\begin{aligned}
&\int\limits_{-\infty}^\infty dx\left[\sin(k'|x|)-k'a_f\cos(k'x)\right]\psi(x,0)\\& \quad\quad\quad\quad\quad\quad
=\frac{2\psi(0,0)}{k'}\left(1-\frac{a_f}{a_i}  \right)+\mathcal{O}\left(\frac{1}{k^2}  \right)
\end{aligned}
\end{equation}
for a wave function whose short range behaves as $\psi(x,0)\approx\psi(0,0)(1-|x|/a_i)$, and whose long range is regular and smooth. This leading-order behavior of the projection encodes the mismatch between the initial and final boundary conditions. Inserting this result into Eq.~\eqref{eq: ball formula 1}, we find that
\begin{equation}\label{eq: ball formula 2}
\tilde{\psi}_{\rm bal}^{(S)}(k,t)= \left(\frac{a_f}{a_i}-1\right)\frac{2\psi(0,0)}{\left(k^2a_f-i|k|\right)}\mathrm{e}^{-iE_k t}+\mathcal{O}\left(\frac{1}{k^3}\right)
\end{equation}
after some algebra. Combining Eqs.~\eqref{eq: sr wave s}, \eqref{eq: sr wave b}, and \eqref{eq: ball formula 2}, we arrive at Eq.~\eqref{eq: ball wave 1d}.


\begin{thebibliography}{99}
\bibitem{endnote3} We have also used this interaction representation to numerically compute the bound state for $a>0$, and we have found that the binding energy and wave function converge to the analytical result in the $\Delta x \ll a$ limit.
\bibitem{endnote4} For quenches to attractive interactions ($a_f>0$), the semiclassical estimate for $P_{CC}$ requires inserting $\tilde{\psi}_{\rm bal}^{(S)}(k,t)$ into Eqs.~\eqref{eq: two body momentum dist}-\eqref{eq: cc estimate} and then adding the probability that the atoms remain bound after the quench. Bound atoms will remain in the central well during the short timeframe of quench-induced transport.
\bibitem{endnote2} More generally, the residues from any poles inside the contour can be absorbed into the definition of $\tilde{\psi}_{\rm bal}^{(S)}(k,t)$. This occurs, for example, if $\psi(x,0)$ already has a ballistic component. Since we are interested in the ballistic response due solely to the interaction quench, we will consider wave functions whose initial large-momentum content comes entirely from the log-derivative boundary condition.
\end{thebibliography}
\end{document}